\documentstyle[12pt,openbib]{article}

\title{{\bf  Renormalization  in Nonrelativistic Quantum Mechanics}
\thanks{To Appear in Journal of Physics A}}

\author {Sadhan K Adhikari\thanks{John Simon Guggenheim Memorial
Foundation Fellow.}
  and  Angsula Ghosh\\ Instituto de F\'\i sica Te\'orica,
Universidade Estadual Paulista,\\ 01405-900 S\~{a}o Paulo, S\~{a}o Paulo,
Brasil \\}

\date{\today } 

\begin{document}
\maketitle

\begin{abstract}

The importance and usefulness of renormalization are emphasized in
nonrelativistic quantum mechanics. The momentum space treatment of both
two-body bound state and scattering problems involving some potentials
singular at the origin exhibits ultraviolet divergence.  The use of
renormalization techniques in these problems leads to finite converged
results for both the exact and perturbative solutions.  The renormalization
procedure 
is carried out for the quantum two-body problem in different partial waves
for a minimal potential possessing only the threshold behavior and no form
factors.  The renormalized  perturbative and exact solutions for this problem
are found to be consistent with each other. The useful role of the
renormalization group equations for this problem is also pointed out.

{\bf PACS Numbers 03.65.-w, 03.65.Nk, 11.10.Gh, 11.10.Hi  }

\end{abstract}

\newpage

\section{Introduction}

The ultraviolet divergences in perturbative quantum field theory can be
eliminated in many cases 
by renormalization to define physical observables, such as charge
or mass, which are often termed the physical scale(s) of the problem
\cite{wil,ryd,ren,we}.  Ultraviolet divergences appear in exact as well as  
perturbative treatments of the nonrelativistic quantum mechanical two-body
problem in momentum space  
interacting via two-body potentials with certain singular behavior at
short distances \cite{wei,ad,adx,jac,tar1,tar2,beg} in two and three space
dimensions.  Renormalization of the potential model leads to a scale(s)
and a finite physical observable(s) \cite{ad}.

Renormalization of a physical quantum mechanical model is essential in
reproducing experimental results irrespective of whether the original model
exhibits ultraviolet divergence or not. Renormalization removes the effect of
different uncertainties and approximations of a physical model on the
observables and brings some of the theoretical predictions in agreement with
experiment. Such uncertainties exist in all quantum mechanical models. Even
in the most well-understood quantum mechanical hydrogen-atom problem only the
long distance behavior of the Hamiltonian can be considered to be known.  For
distances smaller than the radius of the proton, the electron-proton potential
is not the bare Coulomb potential but some regularized Coulomb potential
which, unlike the original Coulomb potential, does not diverge and leads to a
constant value as the electron-proton separation $r$ goes to zero. Also at
this scale the effect of field theoretic corrections to the Hamiltonian is
relevant. The detailed behavior of this regularized Hamiltonian  for small
$r$ depends on the charge distribution of proton and is not usually known.
The role of renormalization is to remove the uncertainty of the regularized
potential by fixing some of the observable(s). The effect of renormalization
in the hydrogen atom problem is small and not evident as the radius of the
proton is very small. The effect is indispensable in large atoms and 
specially in mu-mesic atoms where the orbit of the mu meson could have a
significant overlap with  nuclear matter.

In the above-mentioned Coulomb problem the  potential is divergent at $r =
0$. In spite of the singularity of the Coulomb potential at $r = 0$, both the
scattering and bound state problems with the original potential are solvable
and do not produce ultraviolet divergences.  The role of renormalization 
in this problem is to introduce a regularized well-defined potential which
reproduces some  of the observables.

The situation is different for potentials with a stronger divergence at $r=0$
than the Coulomb potential. These are the potentials   which lead to the 
above-mentioned ultraviolet divergences in momentum space. If these divergent
potentials are attractive at $r=0$,  the original problem does not permit
convergent solution in either momentum or  configuration space. The bound
state problem collapses and produces an infinite number of bound states with
an accumulation point at infinite binding. The scattering Lippmann-Schwinger
equation for these potentials possesses a noncompact kernel and hence is not
amenable to numerical solution.  Finite and meaningful physical solution is
obtained only after renormalization.  If the divergent potentials are
repulsive at $r=0$, in the configuration space treatment one can obtain a
finite convergent solution essentially by imposing some constraints, such as
the solution should vanish at some small $r$. In this way the trouble with
integration over the singular potential near $r=0$ is avoided. In case of
many repulsive divergent potentials, this procedure works and produces
physically meaningful results, some examples being the repulsive soft and
hard core potentials exhibiting  ultraviolet divergences.  Even in these
cases the momentum space treatment, after an appropriate truncation of the
Hamiltonian at small $r$, may lead to a Lippmann-Schwinger
equation with noncompact kernel.  Renormalization is then necessary to produce
finite and physically meaningful results.

The usual difficulty in momentum space treatment with  the Coulomb potential is
the large distance or the infrared divergence. This could be avoided with
the usual Yukawa potential in nuclear and atomic physics.  The Yukawa potential
possesses the same large momentum or short distance behavior as the Coulomb
potential but no infrared divergence. For  a  Yukawa potential without
ultraviolet (and infrared) divergence(s), renormalization improves the large
momentum or short distance convergence properties. In this  problem
renormalization is not necessary but is only desirable. Renormalization makes
this potential smoother and hence easier for numerical and analytic treatment.
However, renormalization is indispensable for problems with ultraviolet
divergence in momentum space.

In this work we shall be limited to the study of renormalization of the three
dimensional two-body problem possessing ultraviolet divergence in momentum
space in close analogy with field theoretic problems. An account of parts of
this work has recently appeared \cite{ad,adx}.
Most of the present
ideas can also be used in two dimensions and in configuration space
treatments \cite{ad,tar1}.  We illustrate the present procedure for a minimal
potential in different partial waves. In momentum space this 
potential possesses only the threshold behavior and is given by $V_L(p',p)= 
p'^L
\lambda p^L$, where $L$ is the angular momentum.  
     As the scattering Lippmann Schwinger
equation has the same generic form for all partial waves, the ultraviolet
divergence of this potential 
model becomes stronger and stronger as $L$ increases.  The leading
ultraviolet divergence of the momentum space integrals, encountered
while solving the
Lippmann-Schwinger equation with this potential, is linear (cubic,...)
in nature for $L=0$ (1,...). The renormalization of this potential
model can be 
 performed by fixing at least one observable, or equivalently, by
introducing at least one physical scale. 
It is also possible to renormalize by introducing more than one physical
scale.  We renormalize both the exact and the perturbative solutions and
find that the renormalized exact and perturbative solutions are consistent
with each other.

We also derive  the renormalization group (RG) equations for this problem.
These equations clearly exhibit the important scaling behaviors of the
different renormalized solutions. The RG equations can be written for
the scattering solutions expressed in terms of certain physical scales
closely related to scattering observables. These equations are valid in
general, independent of the existence of ultraviolet divergence in the
original problem.  Such RG equations and the associated scaling behavior
involving   observables  are  interesting from a physical point of view.

The ultraviolet divergence of the present problem for $L = 0$
can be compared to the
ultraviolet structure  and high energy behavior of the $\lambda\phi^4$ field
theory \cite{ryd,we,tar1,beg}. The super-renormalizable $\lambda\phi^4$ field
theory in 1+1 dimensions 
possesses ultraviolet logarithmic divergence, requires regularization,
and is perturbatively renormalizable \cite{ryd,we}. 
The nonrelativistic scattering problem
with contact interaction in two dimensions also has similar logarithmic 
divergence \cite{tar1}. 
The renormalizable $\lambda\phi^4$ field
theory in 3+1 dimensions has both logarithmic and quadratic divergences
\cite{ryd,we}. 
We have verified that the nonrelativistic scattering problem
with the present minimal potential  in two dimensions also has similar 
logarithmic  and quadratic 
divergences. In the present study of scattering in three dimensions,
although the divergent terms are of different nature, 
the renormalization can be performed in a similar fashion.
 In the field theoretic problem,
 one cannot go beyond few
lowest orders of perturbation theory.  On the other hand, 
the nonrelativistic scattering
problem with the present minimal potential possesses
 stronger ultraviolet divergences than in the $\lambda\phi^4$ field theory
and can be solved to find both the exact and the perturbative solutions
analytically.  In the present  work we find that
the exact renormalized solution is consistent with the perturbative 
one. The study of 
the present analytic model   will allow us to
understand most of  the subtleties of renormalization and RG equations.

The plan of our work is as follows. In Sec. II we perform the
renormalization of the exact solution for the minimal potential in different
partial waves. In Sec. III the renormalization of the perturbative solution
is carried out and consistency of the renormalized exact and perturbative
solutions is demonstrated. In Sec. IV we derive the RG equations and discuss
the scaling properties of the renormalized solution. In Sec. V a brief
summary  of the present work is presented.

\section{Renormalization of the Exact Solution}

The partial-wave Lippmann-Schwinger equation for the scattering amplitude
$T_L(p,q,k^2)$ in three  dimensions, at c.m.  energy $k^2$, is given by
\begin{eqnarray}
T_L(p',p,k^2) = V_L(p',p)+\frac{2}{\pi}\int q^2 dq V_L(p',q)
G(q;k^2)T_L(q,p,k^2),
\label{2} \end{eqnarray} 
with the free Green function $G(q;k^2)=(k^2-q^2+i0)^{-1},$ in units
$\hbar=2m=1$, where $m$ is the reduced mass.

We discuss potential scattering with the minimal  potential in different
partial waves. The present minimal potential in the 
$L$th partial wave is taken
to be $V_L(p', p)= p'^L \lambda p^L$, which is the usual $\delta$ potential
for $L=0$.  
For increasing $L$ this 
potential presents stronger and stronger ultraviolet divergence. 
 The reason for studying this
potential is that it is analytically tractable and presents arbitrarily
strong ultraviolet divergence as $L$ increases. It is not a priori 
clear that
potentials with arbitrarily strong ultraviolet divergence can be
meaningfully  renormalized. Physically, this potential is one of 
arbitrary short range in higher partial waves and should be compared with 
the $S$ wave $\delta$ function potential. If a meaningful solution of the
problem could be found they could be of use in different areas of physics
where the details of a  potential is not of concern. (a)  
The renormalized
solution of the minimal potential could be used in problems of statistical
mechanics, such as, in Cooper pairing in superconductivity.  There are 
evidences of pairing in higher partial waves \cite{ran}.
In this case  the details
of the phonon induced short-range electron-electron potential is irrelevent.
The Cooper and the Bardeen-Cooper-Schrieffer (BCS) equations in
superconductivity,  
have
been satisfactorily 
renormalized in $S$ wave, but not in 
higher partial waves \cite{ran}.  The present work should be of relevance 
to the renormalization of  the Cooper and BCS equations in higher partial 
waves. (b) The present renormalization scheme is also of interest in 
deriving a nucleon-nucleon potential from an effective field theory \cite{eft1}
 as 
suggested by Weinberg \cite{wei}. In this derivation one needs to sum an
infinite series of Feynmann diagrams. In the lowest order, the nucleon-nucleon
potential as derived from the effective field theory includes an attractive
delta potential, the solution of which has been successfully renormalized
\cite{eft1}. However, in higher order one obtains a potential with stronger
divergence involving powers of momentum (see, for example Eq. (3.4) of
Ref. \cite{eft2}). Such a potential should be renormalizable following
the scheme presented here.

The present
approach is also applicable to other potentials with weaker ultraviolet
divergences and/or  permiting  only numerical solution. A numerical study of
the renormalization scheme has recently been made \cite{ad5}.

For the above mentioned minimal potential, the $t$ matrix of Eq. (\ref{2})
permits the following analytic solution
\begin{equation}
T_L(p',p,k^2)=p'^L \tau_L(k) p^L,
\label{4}
\end{equation}
with the $\tau$ function defined by
\begin{eqnarray}
\tau_L(k) & = &  [\lambda^{-1}-I_L(k)]^{-1}, \label{4a}\\
I_L(k)& = & (2/\pi) \int q^2 dq  q^{2L} G(q;k^2).\label{4b}
\end{eqnarray} 
As the $\tau$ function completely determines the $t$ matrix,  we shall
consider only the renormalization of the $\tau$  function.  Here the
condition of unitarity is given by
\begin{equation}
\Im T_L(k) = -k |T_L(k)|^2,\label{uni}
\end{equation}
where $T_L(k)= T_L(k,k,k^2)$ and $\Im$ denotes the imaginary part.

The integral $I_L(k)$ of Eq. (\ref{4b}) possesses ultraviolet divergence. For
$L=0$ (1,...) the leading divergence of this integral is linear (cubic,...)
in nature.  Finite result for the $t$ matrix of Eqs.  (\ref{4}) and
(\ref{4a}) can be obtained only if  $\lambda^{-1}$ also diverges in a similar
fashion and cancels the divergence of $I_L(k)$.  The function  $\lambda
I_L(k)$ is the trace of the kernel of the integral equation (\ref{2}) and
possesses ultraviolet divergence. The kernel of Eq.  (\ref{2}) is noncompact
and it does not have scattering solution.

Hence some regularization is needed to give meaning to Eq.  (\ref{2}). This
can be achieved by using a regularized  Green function involving a cut-off.
One example is the following regularized Green function with a smooth cut-off
$\Lambda$ for $L=0$ as in Ref. \cite{ad}
\begin{eqnarray}
G_R(q,\Lambda;k^2)& = & (k^2-q^2+i0)^{-1}+(\Lambda^2+q^2)^{-1},\nonumber \\ &
= &
\frac{k^2+\Lambda^2}{(k^2-q^2+i0)(\Lambda^2+q^2)}.
\label{9} \end{eqnarray} 
However, in the present work we shall use the following regularized Green
function with a sharp cut-off
\begin{eqnarray}
G_R(q,\Lambda;k^2) = (k^2-q^2+i0)^{-1} \Theta(q-\Lambda), \label{9a}
\end{eqnarray}
$\Theta(x)=0$ for $x > 0$ and =1 for $x < 0$.  In Eqs. (\ref{9}) and
(\ref{9a}) $\Lambda (>>k)$ is a large but finite quantity.  The reason for
choosing Green function (\ref{9a}) is that it is equally applicable for all
types of ultraviolet divergences in all partial waves, whereas Green function
(\ref{9}) is only valid for a linear divergence as encountered in the $L=0$
treatment of Ref. \cite{ad}  and requires modification  if the divergence is
stronger as in this work.  Though we use Green function (\ref{9a}) with the
minimal potential in the present treatment, the present idea of
renormalization can be extended to other (singular) potentials and to other
regularized Green function(s).  The imaginary part of the Green function is
unaffected by this type of regularization, and this  guarantees unitarity
condition (\ref{uni}).

In the end,  the limit $\Lambda \to \infty$ has to be taken, which will
reduce the regularized Green function to the free Green function.  Finite
results for physical magnitudes, as $\Lambda \to \infty$, are obtained only
if the coupling $\lambda$ is also replaced by the so called bare coupling
$\lambda_L(k,\Lambda)$. The choice of the bare coupling is different for
different $L$ and can be found by inspection of the following regularized
expressions of the integral $I_L(k)$ of (\ref{4b}) for different $L$
\begin{eqnarray}
I_{RL}(k,\Lambda) & \equiv & \frac{2}{\pi}\int q^2 dq q^{2L}
G_R(q,\Lambda;k^2),\\ & = &
-\frac{2}{\pi}\biggr[\Lambda+\frac{k}{2}\ln|\frac{\Lambda-k}{
\Lambda+k}| \biggr]-ik, L=0, \\
  & =& -\frac{2}{\pi}\biggr[\frac{\Lambda^3}{3}+k^2\Lambda
+\frac{k^3}{2}\ln|\frac{\Lambda-k}{
\Lambda+k}| 
 \biggr]-ik^3, L = 1,\\ & =& -\frac{2}{\pi}\biggr[\frac{\Lambda
^5}{5}+\frac{k^2\Lambda^3}{3} +k^4\Lambda
+\frac{k^5}{2}\ln|\frac{\Lambda-k}{
\Lambda+k}|    \biggr] -ik^5, L=2.
\end{eqnarray}
Consistent with the  large $\Lambda (>> k)$ limit, the logarithmic terms in
the above expressions for $I_{RL}(k,\Lambda)$ tend to zero, and for a general
$L$ we have in this limit
\begin{eqnarray}
I_{RL}(k,\Lambda)=-
\frac{2}{\pi} \sum_{i=0}^L  \frac{k^{2(L-i)}\Lambda^{2i+1}}{2i+1}
 -ik^{2L+1}.\label{101}
\end{eqnarray}
All the terms in the summation in Eq. (\ref{101}) diverges as $\Lambda \to
\infty.$ Except for $L=0$, these divergent terms are momentum $(k)$
dependent. In the present work, the leading divergence is much stronger for
a general $L$ compared to the $S$ wave case. In Eq. (\ref{101}) the leading
divergence is like $\Lambda^{2L+1}$.  The $S$ wave treatment of the $\delta$
potential in Ref. \cite{ad} had only an energy independent term diverging
linearly as $\Lambda$. For a finite $k (<<\Lambda)$, the stronger divergence
and the energy dependence of the divergent terms in the present case do not
introduce any complication and the ideas of Ref. \cite{ad} can be generalized.

In order to obtain a finite renormalized $\tau$ function, the coupling
$\lambda$ should be replaced by the so called bare coupling  defined, for
example,  by
\begin{eqnarray}
\lambda_{L}^{-1}(k,\Lambda) & = & -\frac{2}{\pi} \sum_{i=0}^L
\frac{k^{2(L-i)}\Lambda^{2i+1}}{2i+1} - \Lambda_{0L}(k^2), \label{x}
\end{eqnarray}
where the function $\Lambda_{0L}(k^2)$   defines the physical scale(s) of the
system and characterizes the interaction.  In the end the
physical scale(s) in  $\Lambda_{0L}(k^2)$  should be identified with a
physical observable(s). If the problem is characterized by a single 
physical scale,
e.g., the scattering length $a_L$,  it is appropriate to take
$\Lambda_{0L}(k^2)$ to be independent of $k^2$: $\Lambda_{0L}(k^2) =
-1/a_L^{2L+1}$.  If the problem is characterized by two physical 
scales, such as a
scattering length $a_L$ and another physical 
scale  $b_L$, it is natural to take the
following  expansion
\begin{equation}
\Lambda_{0L}(k^2) =-
1/a_L^{2L+1}- b_L^{1-2L} k^2.
\end{equation} 
We have taken both the scales $a_L$ and $b_L$ to have the dimension of
length. A third scale $c_L$ can be accommodated similarly through
\begin{equation}
\Lambda_{0L}(k^2) =
-1/a_L^{2L+1}- b_L^{1-2L} k^2 -c_L^{3-2L}  k^4,\label{xsca}
\end{equation} 
where $c_L$  has also been chosen to have the dimension of length. Equation
(\ref{xsca}) is just a Taylor series expansion of $\Lambda_{0L}(k^2)$ at low
energies. It is realized that in the present renormalization 
 the number of divergent terms 
and the number of scales are not related.

The regularized $\tau$ function of Eq. (\ref{4a}) can now be rewritten as
\begin{eqnarray}
\tau_{L}(k,\Lambda)& = &[\lambda_L^{-1}(k,\Lambda)-I_{RL}(k,\Lambda)]^{-1} ,
\label{8} \end{eqnarray}
  where  for a finite $\Lambda$, $I_{RL}(k,\Lambda) $ is a convergent
integral.  As $\Lambda \to
\infty$, however, this integral develops the original ultraviolet divergence.
In this limit, the quantity $\lambda_L^{-1}(\Lambda,k)$ of Eq.  (\ref{x}) has
the appropriate divergent behavior,  that cancels the divergent parts of
$I_{RL}(k,\Lambda)$.  In Eq. (\ref{8}) the explicit dependence of the $\tau$
function on  $\Lambda$ has been introduced.

Next the limit $\Lambda \to \infty$ has to be taken in Eq. (\ref{8}).  With
this regularization, the renormalized $\tau$ function can be written as
\begin{eqnarray}
\tau_{RL}(k,\lambda_R(k,\mu),\mu)& = &[\lambda^{-1}_{RL}
(k,\mu)-I_{RL}(k,\mu)]^{-1} ,
\label{200} \end{eqnarray}
where $\mu$  is  the scale of the problem and emerges as a result of
renormalization. The renormalization scale $\mu$ should be contrasted with
the physical scale(s) in  $\Lambda_{0L}(k^2)$.  The renormalized $\tau$
function will be independent of $\mu$. In Eq. (\ref{200})  the explicit
dependence of the $\tau$ function on both $\mu$ and the renormalized coupling
$\lambda_{RL}(k,\mu)$ has been exhibited.  The limiting procedure implied by
$\Lambda \to \infty$ in Eq. (\ref{8}) leads to the following definition for
the renormalized coupling $\lambda_{RL}(k,\mu)$
\begin{eqnarray}
\lambda_{RL}^{-1}(k,\mu)   = \lim_{\Lambda \to \infty}
[\lambda_L^{-1}(k,\Lambda) -\{I_{RL}(k,\Lambda)-I_{RL}(k,\mu)\}].
\label{207} \end{eqnarray}
In Eq. (\ref{207}), if  the  limit $\Lambda \to \infty$ taken, we get
\begin{equation}
\lambda_{RL}(k,\mu) =  \lambda_L(k,\Lambda =\mu). \label{15}\end{equation}
Relation  (\ref{15})
 between the renormalized coupling and bare coupling depends on
the regularization scheme used.  Equations (\ref{x})  and (\ref{15}) lead to
the following expression for the renormalized coupling
\begin{eqnarray}
\lambda^{-1}_{RL}(k,\mu) & = & 
 -\frac{2}{\pi} \sum_{i=0}^L
\frac{k^{2(L-i)}\mu^{2i+1}}{2i+1} - \Lambda_{0L}(k^2). \label{x1}
\end{eqnarray}
The renormalized coupling for two renormalization 
scales $\mu$ and $\mu_0$ are related by the
following flow equation
\begin{equation}
\lambda^{-1}_{RL}(k,\mu)  
 +\frac{2}{\pi} \sum_{i=0}^L
\frac{k^{2(L-i)}\mu^{2i+1}}{2i+1} =
\lambda^{-1}_{RL}(k,\mu_0) +
 \frac{2}{\pi} \sum_{i=0}^L
\frac{k^{2(L-i)}\mu_0^{2i+1}}{2i+1}. \label{x2}
\end{equation}
For a general $L$,
this flow equation is energy dependent but independent of the regularization
scheme.  For $L=0$, as in Ref. \cite{ad}, the renormalized coupling and the 
flow equations are  energy independent. The absolute value of the 
renormalized coupling $\lambda_{RL}(k,\mu)$ increases with
$\mu$.  Thus if we start with a small $\lambda_{RL}(k,\mu_0)$ at a given
renormalization 
scale $\mu_0$, the effective coupling constant increases with $\mu$ as in the
$\lambda \phi^4$ model
\cite{ryd}. With the increase of $\mu$  one can reach a large enough
$\lambda_{RL}(k,\mu)$, where perturbative treatment is not valid.  The energy
dependence of the renormalized coupling (\ref{x1}) and the flow equation
(\ref{x2}) for $L\ne 0$ does not create any complication and one can
renormalize the results and write the RG equations.

The present scattering model  permits analytic solutions for all $L$.  The
renormalized $\tau$  function is given by
\begin{equation}
\tau_{RL}(k,\lambda_{RL}(k,\mu),\mu) =  \biggr[\lambda_{RL}^{-1}(k,\mu)+
\frac{2}{\pi} \sum_{i=0}^L
\frac{k^{2(L-i)}\mu^{2i+1}}{2i+1}
+ik^{2L+1}\biggr]^{-1},
\label{307}\end{equation}
 Explicitly, using  the renormalized coupling  (\ref{x1}), the renormalized
$\tau$ function can be written as
\begin{equation}
\tau_{RL}(k,\lambda_{RL}(k,\mu),\mu) =  [ik^{2L+1}-\Lambda_{0L}(k^2)]
^{-1}. \label{ex}
\end{equation}
This $\tau$ function depends on the renormalized coupling $\lambda_{RL}
(k,\mu)$, but not on $\mu$, that is the explicit and implicit (through
$\lambda_{RL}(k,\mu)$) dependences of the $\tau$ function  on $\mu$ cancel.
Physics is determined by the value of $\lambda_{RL} (k,\mu)$ at an arbitrary
value of $\mu$ \cite{tar1}, or the following $\mu$ independent quantity
\begin{equation}
\lambda_{RL}^{-1}(k,\mu)+
 \frac{2}{\pi} \sum_{i=0}^L
\frac{k^{2(L-i)}\mu^{2i+1}}{2i+1}
= - \Lambda_{0L}(k^2), \end{equation} 
as can be seen from Eqs. (\ref{x1}) and (\ref{307}).

From Eq. (\ref{ex}) we find that the renormalized $\tau$ function is a
function of $\Lambda_{0L}(k^2)$. It is convenient to express the renormalized
$\tau$ function in terms of the physical scales $a_L$, $b_L$, and $c_L$
introduced in Eq. (\ref{xsca}).  Once this is done, $\tau_{RL}$ is determined
by the physical scale(s) which are closely related to the observables of the
system.  Then  Eq. (\ref{ex}) reduces to
\begin{eqnarray}
\tau_{RL}(k,a_L,b_L,c_L)\equiv
\tau_{RL}(k,\lambda_R(k,\mu),\mu) = 
[ik^{2L+1}+1/a_L^{2L+1}+b_L^{1-2L} k^2  +c_L^{3-2L}  k^4]^{-1}.
\label{301} \end{eqnarray} 
The name `physical 
scale' given to $a_L$, $b_L$, and $c_L$, is justified as these
quantities are  a measure of low energy scattering in each partial wave.

We have here renormalized a divergent physical problem and obtained the
well-defined 
solution (\ref{301}). As the original problem is ill-defined, it is 
interesting to ask if this renormalized solution is physically acceptable
or is just a finite answer obtained by a mathematical trick from the 
unregularized original problem. The fact, that the renormalized result 
is physically motivated, can now be established by a careful examination. 
For $L=0$ Eq. (\ref{301}) is just the usual effective range expansion for 
the $t$ matrix \cite{rod}. 
The same is also true for higher partial waves \cite{rod}. Hence the 
renormalized solution (\ref{301}) is the physically expected solution of
the problem for a short-range potential and should lead to acceptable results
for other observables. So the present renormalization scheme for $L\ne 0$
should be considered as a natural generalization of our previous
results presented in Ref. \cite{ad}. 

To bring further evidence to the acceptability of the present result and 
to demonstrate  the self-consistency of the present renormalization scheme,
we perform perturbative renormalization of the same problem in the next 
section and establish the equivalence between the two approaches.

\section{Perturbative Renormalization}

In the last section we performed the  renormalization of the exact analytic
solution. In the simplest field theoretic $\lambda \phi^4$ model the exact
solution is not known because of the creation and annihilation of particles
and also because of the quartic nature of the interaction. In that case, 
 one usually performs perturbative renormalization. Though it is expected
that the result of perturbative renormalization should be consistent with
that of exact renormalization, there is no general proof in this regard.  As
the present problem is much simpler than the 
$\lambda \phi ^4$ model, it is also
illustrative to perform perturbative renormalization of the present problem
and to show that the result is consistent with the exact renormalization of
the last section.  This consistency can also  be established in the 
case of the exactly soluble Schwinger model for massless quantum 
electrodynamics in 1+1
dimensions.

From Eq. (\ref{4a}) the perturbative solution of the present problem is given
by
\begin{equation}
\tau_L(k)= \lambda [1 + \lambda I_{RL}(k,\Lambda)+
 \lambda^2 I_{RL}^2(k,
\Lambda)
+\lambda^3  I^3_{RL}(k,\Lambda) +... ], \label{13}
\end{equation}
where we have used the regularized version of the integral
$I_{RL}(k,\Lambda)$ given by (\ref{101}).

Up to the first order in perturbation theory in Eq. (\ref{13}), we take
$\lambda_L^{(1)} = \bar \lambda_L$ and consequently to this order in the
redefined coupling strength $\bar \lambda_L$, $\tau_{L}^{(1)}(k) = \bar
\lambda_L$ and there is no divergence as $\Lambda
\to \infty$. In order to find the result finite to first order one could 
have taken $\lambda_L^{(1)} =  \lambda$. But if we would like  to obtain finite
results in all orders of perturbation theory, which are consistent with the
exact renormalized result of the last section, a different finite coupling
$(\bar \lambda_L)$ has to be introduced in all orders as in this section.

The second order $\tau$ function with this $\bar \lambda_L$, given by,
\begin{equation}
\tau_{L}^{(2)}(k)=\bar \lambda_L
 [1 + \bar\lambda_L I_{RL}(k,\Lambda)], \label{14}
\end{equation}
however, diverges as $\Lambda \to \infty$, because in this limit the
regularized integral $I_{RL}(k,\Lambda)$ of (\ref{101}) diverges. A finite
result for the $\tau$ function up to the second order in $\bar\lambda_L$  could
be obtained by employing  the following coupling
\begin{equation}
\lambda_L^{(2)} =\bar \lambda_L \biggr[1+\bar \lambda_L \frac{2}{\pi} \sum_{i=0}^L
\frac{k^{2(L-i)}\Lambda^{2i+1}}{2i+1}   \biggr]. \label{24a}
\end{equation}
With this modified coupling, the second order $\tau$ function (\ref{14}) is
given by
\begin{equation}
\tau_{L}^{(2)}(k)=\lambda_L^{(2)} [1 + \lambda_L^{(2)} 
I_{RL}(k,\Lambda)], \label{sec}
\end{equation}
where $I_{RL}(k,\Lambda)$ is given by Eq. (\ref{101}). With $\lambda_L^{(2)}$
given by (\ref{24a}), the second order $\tau$ function, $\tau_{L}^{(2)}(k)$,
contains terms up to the 
fourth order in $\bar \lambda_L$. Up to the second
order in $\bar
\lambda_L$,  $\tau_{L}^{(2)}(k)$
is finite in the limit $\Lambda
\to \infty$ and is given by 
\begin{equation}
\tau_{L}^{(2)}(k)= \bar\lambda_L [1 - \bar\lambda_L ik^{2L+1}]. \label{16}
\end{equation}

With the second order coupling constant  given by Eq. (\ref{24a}), the third
order $\tau$ function
\begin{equation}
\tau_{L}^{(3)}(k)= \lambda_L^{(2)} [1 + \lambda_L^{(2)} I_{RL}(k,\Lambda)
+ \lambda_L^{(2)^2}  I^2_{RL}(k,\Lambda) ] \label{20}
\end{equation}
diverges in the limit  $\Lambda \to \infty$. In order to obtain a finite
$\tau$ function in this limit up to the third order in $\bar \lambda_L$, one
 could employ the following third order $\lambda$
\begin{equation}
\lambda_L^{(3)} = 
\bar\lambda_L \biggr[1+ \bar\lambda_L \frac{2}{\pi} \sum_{i=0}^L
\frac{k^{2(L-i)}\Lambda^{2i+1}}{2i+1}+
\bar\lambda_L^2 \biggr( \frac{2}{\pi} \sum_{i=0}^L
\frac{k^{2(L-i)}\Lambda^{2i+1}}{2i+1}\biggr)^2   \biggr] \label{14a}
\end{equation}
in  the following expression for the third order $\tau $ function
\begin{equation}
\tau_{L}^{(3)}(k)= \lambda_L^{(3)} [1 + \lambda_L^{(3)} I_{RL}(k,\Lambda)
+ \lambda_L^{(3)^2}  I^2_{RL}(k,\Lambda) ].
\end{equation}
 This third order $\tau$ function now contains terms up to the sixth order in
$\bar \lambda_L$. If the third order $\tau$ function is truncated up to 
third order terms in $\bar\lambda_L $, and the limit $\Lambda \to \infty$ is
taken, we obtain
\begin{equation}
\tau_{L}^{(3)}(k)=\bar \lambda_L [1 - \bar\lambda_L ik^ {2L+1}
+ \bar\lambda_L^2 (-ik^{2L+1})^2]. \label{30}
\end{equation}

With the third order $\lambda$ given by Eq. (\ref{14a}), the higher order
$\tau$ functions $\tau_{L}^{(l)}, l>3$ diverges in the limit $\Lambda \to
\infty$.  In order to obtain a finite $\tau_{L}^{(l)}, l>3$, one should modify
the coupling strength $\lambda$.   A finite $\tau_{L} ^{(l)}$ up to $l$th
order, in the limit $\Lambda \to \infty$, can be obtained by employing  the
following coupling
\begin{equation}
\lambda_L^{(l)} = \bar\lambda_L \biggr[1+ \sum_{j=1}^{l-1}
 \biggr( \bar\lambda_L\frac{2}{\pi} \sum_{i=0}^L
\frac{k^{2(L-i)}\Lambda^{2i+1}}{2i+1}\biggr)^j   \biggr]. \label{34a}
\end{equation}
With this $\lambda$, the $l$th order $\tau$ function is given by
\begin{equation}
\tau_{L}^{(l)}(k)= \lambda_L^{(l)} [1 +\sum_{j=1}^{l-1}
 \lambda_L^{(l)^j}  I^j_{RL}(k,\Lambda)]. \label{21}
\end{equation}
Once the limit $\Lambda \to \infty$ is taken in Eq. (\ref{21}) and terms up
to the order of $\bar\lambda_L^l $ are maintained,  the following
result is obtained
\begin{equation}
\tau_{L}^{(l)}(k)= \bar\lambda_L 
[1 + \sum_{j=1}^l(-\bar \lambda_L ik^ {2L+1})^j].
\label{31}
\end{equation}
Unlike in the case of $\lambda \phi^4$ field theory, one can calculate the
result to an arbitrarily large order in perturbation theory.  The summation
in Eq. (\ref{31}) is a geometric series and as $l \to \infty$ this series can
be summed to yield
\begin{equation}
\tau_{RL}(k) = \frac{1}{1/\bar\lambda_L +ik^{2L+1}}, \label{25}
\end{equation}
which is the result of perturbative renormalization.

If we compare the result of perturbative renormalization (\ref{25}) with the
exact renormalized solution (\ref{ex}) we realize that these two are
equivalent if $\Lambda_{0L}(k^2)$ of Eq. (\ref{ex}) is identified as $-\bar
\lambda_L^{-1}$
of Eq. (\ref{25}). If this identification is made, the result of perturbative
renormalization is consistent with the exact renormalized result.
Then we find that the parameter $\Lambda_{0L}(k^2)$ 
of exact renormalization is intimately
related to the strength parameter $\bar \lambda_L$ of perturbative
renormalization.

\section{Renormalization Group Equations}

The renormalized $\tau$ function is independent of $\mu$, so is invariant under
the group of transformations $\mu  \to  {\exp}(s)\mu$, which form the
RG.  In the present case, as in the $\lambda \phi^4$
model, it is convenient to work in terms of the dimensionless coupling,
$g_{RL}(\mu)$, defined by
\begin{eqnarray}
g_{RL}(k,\mu)  \equiv  \mu^{2L+1} \lambda_{RL}(k,\mu), 
 \label{355}  
\end{eqnarray} 
The renormalization condition is given by
\begin{equation} 
\mu \frac{d}{d\mu} \tau_{RL}(k,g_{RL}(k,\mu),\mu) =0,
\label{360}\end{equation}
or,
\begin{equation}
\left[ \mu \frac{\partial}{\partial \mu}+ \beta_L(k,g_{RL}(k,\mu),\mu)
\frac{\partial}
{\partial g_{RL}}
\right]\tau_{RL}(k,g_{RL}(k,\mu),\mu)=0,
\label{370}\end{equation}
where
\begin{equation}
\beta_L(k,g_{RL}(k,\mu),\mu)=\mu \frac{\partial g_{RL}(k,\mu)}{\partial \mu}.
\label{380}\end{equation}
Equation (\ref{370}) is the RG equation.

As the present problem permits analytic solution, the constant
$\beta_L$ of Eq. (\ref{380}) can be exactly calculated. 
From Eqs.  (\ref{355}) and
(\ref{380}) we have
\begin{equation}
\beta_L(k,g_{RL}(k,\mu),\mu) =(2L+1){g_{RL}(k,\mu)}+
\mu^{2L+2} \frac{\partial
\lambda_{RL}(k,\mu)} {\partial \mu}.
\label{440} \end{equation}
With $\lambda_{RL}(k,\mu)$ defined by Eq. (\ref{x1}), we have
from Eqs. (\ref{355}) and  (\ref{440})
\begin{equation}
\beta_L(k,g_{RL}(k,\mu),\mu) = (2L+1)g_{RL} + g_{RL}^2 \frac{2}{\pi}
\sum_{i=0}^L{k^{2(L-i)}\mu^{2(i-L)}}. 
\label{460} \end{equation}
For $L=0$
the $\beta$ function is energy independent and depends implicitly 
on $\mu$ through coupling $g_{RL}$, whereas for $L \ne 0$
the $\beta$ function has explicit dependence on both energy and $\mu$.

The following equation  expresses the invariance of the $\tau$
function  $\tau_{RL}(k,g_{RL}(k,\mu),\mu)$ under a change of momentum scale:
\begin{equation}
\tau_{RL}
(\gamma k,g_{RL}(k,\mu),\mu)=
\gamma^{-(2L+1)}\tau_{RL}(k,g_{RL}(k,\mu),\mu
\gamma^{-1}).\label{sca1}
\end{equation}
Equations (\ref{307}) and (\ref{355}) are consistent with scaling 
(\ref{sca1}).
In Eq. (\ref{sca1}) the change of scale is effected on the explicit 
momentum ($k$) dependence of the $\tau$ function and not on the implicit 
momentum dependence of the coupling constant $g_{RL}$.
From Eq. (\ref{sca1})  we obtain
\begin{equation}
\left[\gamma\frac{\partial}{\partial \gamma}+\mu\frac{\partial }{\partial
\mu} +(2L+1) \right] \tau_{RL}(\gamma k,g_{RL}(k,\mu),\mu) = 0.
\label{390} \end{equation}
Eliminating the partial derivative $\mu(\partial \tau_{RL}
/\partial \mu)$ between
Eqs. (\ref{370})  and (\ref{390}) we have
\begin{equation}
\left[ \gamma\frac{\partial}{\partial
\gamma}-\beta_L(k,g_{RL}(k,\mu),\mu)
\frac{\partial}{\partial g_{RL}}+(2L+1)   \right]
\tau_{RL}(\gamma k,g_{RL}(k,\mu),\mu) = 0,  \label{410}
\end{equation}
with $\beta_L$ given by 
Eq. (\ref{460}).  RG equation (\ref{410}) expresses the effect on the
$\tau$ function  of scaling up momentum by a factor $\gamma$.

The RG equations (\ref{370}) and (\ref{410}) involve the renormalized
coupling $g_{RL}$ and the renormalization scale $\mu$ and  are not closely
related to the physical observables. However, one can write equivalent RG
equations in terms of the physical scales $a_L$, $b_L$, and $c_L$
of Eq. (\ref{301}), which are closely related to experimental 
observables. From Eqs.  (\ref{307}), (\ref{301}), and 
(\ref{355})  one has the identity
\begin{equation}
\left[ \gamma\frac{\partial}{\partial
\gamma}-\beta_L
\frac{\partial}{\partial g_{RL}}   \right]
\tau_{RL}(\gamma k,g_{RL},\mu) =
 \left[ \gamma\frac{\partial}{\partial
\gamma}-a_L\frac{\partial}{\partial a_L}- 
b_L\frac{\partial}{\partial b_L}-c_L\frac{\partial}{\partial c_L}
  \right]  \tau_{RL}(\gamma
k,a_L,b_L,c_L) , 
\end{equation}
so that the RG equation (\ref{410}) becomes
\begin{equation}
 \left[ \gamma\frac{\partial}{\partial
\gamma}-a_L\frac{\partial}{\partial a_L}-
b_L\frac{\partial}{\partial b_L}-c_L\frac{\partial}{\partial c_L}
+(2L+1)   \right]  \tau_{RL}
(\gamma
k,a_L,b_L,c_L) = 0,  \label{29}
\end{equation}
Equations (\ref{410}) and  (\ref{29})
express the fact that the effect of a change in the momentum scale 
$\gamma$ on $\tau_{RL}$ can be
compensated by the effect of a change in $g_{RL}$ or equivalently, 
in $ a_L$, $b_L$, and $c_L$,  respectively.
In RG equation  (\ref{29})  $a_L$, $b_L$, and $c_L$  are physical scales. 
RG equation (\ref{29}) implies the following scaling
\begin{equation}
\tau_{RL}
(\gamma
k,a_L,b_L,c_L)=\gamma^{-(2L+1)}\tau_{RL}
(k,\gamma a_L,\gamma b_L, \gamma c_L). \label{39}
\end{equation}
Hence from the knowledge of the $\tau$ function or the $t$ matrix at a
certain energy one can predict the $\tau$ function at another energy. 
RG equations allow one to extrapolate the $\tau$ function from
one energy to another.

In principle, RG equations  can  be solved to yield the exact
renormalized $\tau$ function. However, it is illustrative  to obtain the
asymptotic high-energy behavior  of this $\tau$ function  from  RG equations
(\ref{410}) or (\ref{29}).  At high energies $\gamma \to \infty$, and
 Eqs.  (\ref{410}) or  (\ref{29})
reduces to
\begin{equation}
\gamma\frac{{\partial} \tau_{RL}(\gamma k)}{{\partial} \gamma}+(2L+1)
\tau_{RL}(\gamma k)=0.
\label{40} \end{equation}
This has the simple solution $\lim_{\gamma\to \infty}\tau_{RL}(\gamma k) \sim
1/\gamma^{2L+1}$ again consistent  with the $\tau$ function of Eq. (\ref{301}).

The RG equations of this section 
yield certain general scaling properties  of the renormalized $\tau$ function.
Similar RG  equations  should be valid in general for potentials with certain
renormalizable singular behavior at short distances.  The RG equations in
terms of the physical scales and the associated scaling relations, e.g. Eqs.
 (\ref{29}), (\ref{39}), and (\ref{40}), 
should be valid in general independent of whether the original problem had
ultraviolet divergence or not. Hence such equation should be useful in
general. Obviously, such equation could now be generalized to incorporate
more physical scales.

\section{Summary}

We have emphasized the role of renormalization in nonrelativistic quantum 
mechanics. Renormalization is desirable in most of quantum mechanical bound
state and scattering problems, if one is interested in comparing the result
of a physical model with  experimental observables. Renormalization is
essential in some problems exhibiting ultraviolet divergence, as in quantum
field theory, in order to yield well-defined and finite observables. In both 
cases the final renormalized results could be expressed in terms of certain
physical scales which are closely related to physical observables. 

We have renormalized a potential model exhibiting 
ultraviolet divergence in all partial waves. 
As this model permits analytic solution, we have
renormalized the solution exactly and also perturbatively. In field theoretic
model only perturbative renormalization is possible. As the present model 
permits both perturbative and exact solutions it gives us the unique 
opportunity to test the equivalence of the two.  Such equivalence is
established under very general conditions. The final renormalized result can
be expressed equivalently, in terms of a renormalized coupling $\lambda_{RL}$
and  renormalization scale $\mu$, or in terms of some physical scales related
to observables. 

Finally, we derived RG equations for the renormalized amplitudes expressed in
terms of both renormalized coupling and physical scales. Though the physical 
content of both are identical, RG equations in terms of physical scales 
seem to be more useful from a practical point of view. Such RG equations in
terms of physical scales are valid irrespective of the existence 
of the ultraviolet divergence in the original equation. These equations 
provide interesting scaling behavior of the physical scattering amplitude. 
The RG equations
are expected to be very useful in situations where the analytic solution is
not known, for example, in other few- and many-body problems. The study of 
renormalization and RG
equations in these cases will be an interesting topic for future
investigation.

We thank Dr. R. Banerjee, Dr. T. Frederico, Dr. B. M. Pimentel,  and Dr. 
L. Tomio for discussions
and the Conselho Nacional de Desenvolvimento Cient\'{\i}fico e Tecnol\'ogico,
Funda\c c\~ao de Amparo \`a Pesquisa do Estado de S\~ao Paulo, and
Financiadora de Estudos e Projetos
of Brazil for partial financial support.

\end{document}